\documentclass[reprint,prc,aps,superscriptaddress,floatfix,showpacs,showkeys,altaffilletter]{revtex4-1}

\newcommand{\nuc}[2]{$^{#1}$#2}
\newcommand{\jp}[3]{$J^{\pi}$~=~#1$^{#2}_{#3}$}

\usepackage{comment}
\usepackage[pdftex]{graphics}      
\usepackage{graphicx}                  
\usepackage[rightcaption]{sidecap}
\usepackage{epsfig}
\usepackage{bm}
\usepackage{verbatim}   
\usepackage{color}         
\usepackage[percent]{overpic}
\usepackage{multirow}
\usepackage{longtable}
\usepackage{tablefootnote}
\usepackage{footnote}
\usepackage{dcolumn}
\usepackage{setspace}
\usepackage{tabularx}
\usepackage{tablefootnote}
\usepackage{array}
\usepackage[T1]{fontenc}
\usepackage{epstopdf}
\usepackage{hyperref}
\hypersetup{
     colorlinks   = true,
     citecolor    = blue,
     urlcolor     = blue,
     linkcolor   = blue,
}

\begin{document}


\title{Ground-state and decay properties of neutron-rich \nuc{\bf{106}}{Nb}}



\author{A.~J.~Mitchell}
\email[Email: ]{aj.mitchell@anu.edu.au}
\affiliation{Department of Nuclear Physics, Research School of Physics, The Australian National University, Canberra, ACT 2601, Australia}

\author{R.~Orford}
\altaffiliation[Present address: ]{Nuclear Science Division, Lawrence Berkeley National Laboratory, Berkeley, California 94720, USA.}
\affiliation{Department of Physics, McGill University, Montreal, Quebec H3A 2T8, Canada}
\affiliation{Physics Division, Argonne National Laboratory, Argonne, IL 60439}

\author{G.~J.~Lane}
\affiliation{Department of Nuclear Physics, Research School of Physics, The Australian National University, Canberra, ACT 2601, Australia}

\author{C.~J.~Lister}
\affiliation{Department of Physics and Applied Physics, University of Massachusetts Lowell, Lowell, MA 01854}

\author{P.~Copp}
\altaffiliation[Present address: ]{Physics Division, Argonne National Laboratory, Argonne, IL 60439}
\affiliation{Department of Physics and Applied Physics, University of Massachusetts Lowell, Lowell, MA 01854}

\author{J.~A.~Clark}
\affiliation{Physics Division, Argonne National Laboratory, Argonne, IL 60439}

\author{G.~Savard}
\affiliation{Physics Division, Argonne National Laboratory, Argonne, IL 60439}
\affiliation{Department of Physics, University of Chicago, Chicago, IL 60637}

\author{J.~M.~Allmond}
\affiliation{Physics Division, Oak Ridge National Laboratory, Oak Ridge, TN 37830}

\author{A.~D.~Ayangeakaa}
\affiliation{Department of Physics and Astronomy, University of North Carolina at Chapel Hill, Chapel Hill, North Carolina 27599-3255}
\affiliation{Triangle Universities Nuclear Laboratory, Duke University, Durham, North Carolina 27708-2308}

\author{S.~Bottoni}
\altaffiliation[Present address: ]{Universita' degli Studi di Milano, Instituto Nazionale di Fisica Nucleare, Milano 20133, Italy}
\affiliation{Physics Division, Argonne National Laboratory, Argonne, IL 60439}

\author{M.~P.~Carpenter}
\affiliation{Physics Division, Argonne National Laboratory, Argonne, IL 60439}

\author{P.~Chowdhury}
\affiliation{Department of Physics and Applied Physics, University of Massachusetts Lowell, Lowell, MA 01854}

\author{D.~A.~Gorelov}
\affiliation{Physics Division, Argonne National Laboratory, Argonne, IL 60439}
\affiliation{Department of Physics and Astronomy, University of Manitoba, Winnipeg, Manitoba R3T 2N2, Canada}

\author{R.~V.~F. Janssens}
\affiliation{Department of Physics and Astronomy, University of North Carolina at Chapel Hill, Chapel Hill, North Carolina 27599-3255}
\affiliation{Triangle Universities Nuclear Laboratory, Duke University, Durham, North Carolina 27708-2308}

\author{F.~G.~Kondev}
\affiliation{Physics Division, Argonne National Laboratory, Argonne, IL 60439}

\author{U.~Patel}
\altaffiliation[Present address: ]{Department of Physics, Duke University, Durham, North Carolina 27701}
\affiliation{Department of Nuclear Physics, Research School of Physics, The Australian National University, Canberra, ACT 2601, Australia}

\author{D.~Seweryniak}
\affiliation{Physics Division, Argonne National Laboratory, Argonne, IL 60439}

\author{M.~L.~Smith}
\altaffiliation[Present address: ]{Australian Nuclear Science and Technology Organisation, Lucas Heights, NSW 2234, Australia}
\affiliation{Department of Nuclear Physics, Research School of Physics, The Australian National University, Canberra, ACT 2601, Australia}

\author{Y.~Y.~Zhong}
\affiliation{Department of Nuclear Physics, Research School of Physics, The Australian National University, Canberra, ACT 2601, Australia}

\author{S.~Zhu}
\altaffiliation[Present address: ]{Brookhaven National Laboratory, Upton, NY 11973}
\affiliation{Physics Division, Argonne National Laboratory, Argonne, IL 60439}

\date{\today}


\begin{abstract}

The ground-state properties of neutron-rich \nuc{106}{Nb} and its $\beta$ decay into \nuc{106}{Mo} have been studied using the CARIBU radioactive-ion-beam facility at Argonne National Laboratory. Niobium-106 ions were extracted from a \nuc{252}{Cf} fission source and mass separated before being delivered as low-energy beams to the Canadian Penning Trap, as well as the X-Array and SATURN $\beta$-decay-spectroscopy station. The measured \nuc{106}{Nb} ground-state mass excess of -66202.0(13)~keV is consistent with a recent measurement but has three times better precision; this work also rules out the existence of a second long-lived, $\beta$-decaying state in \nuc{106}{Nb} above 5~keV in excitation energy. The decay half-life of \nuc{106}{Nb} was measured to be 1.097(21)~s, which is 8$\%$ longer than the adopted value. The level scheme of the decay progeny, \nuc{106}{Mo}, has been expanded up to $\approx$4~MeV. The distribution of decay strength and considerable population of excited states in \nuc{106}{Mo} of $J\geq3$ emphasises the need to revise the adopted \jp{1}{-}{} ground-state spin-parity assignment of \nuc{106}{Nb}; it is more likely to be $J\geq3$. 
\end{abstract}


\maketitle


\section{INTRODUCTION}\label{sec:introduction}

Atomic nuclei that bridge the chart of nuclides between the so-called `valley of stability' and `neutron drip-line' play diverse roles in nuclear science. As well as providing important tests of fundamental nuclear-structure theory, quantitative measurements of their ground-state and decay properties provide highly valued constraints of stellar nucleosynthesis models \cite{b2hf} and decay-heat calculations for the nuclear energy sector \cite{iaea}. 

The flow of $r$-process nucleosynthesis across the neutron-rich landscape is largely dictated by the near-parabolic shape of the valley of stability. Variations in binding energy per nucleon along isobaric chains determine both the extreme limit of the neutron drip-line and each nuclide's $Q$-value for $\beta$ decay back towards stability, thereby modulating the timescale of the entire process. To a large extent, this parabolic shape is a result of the bulk properties of nuclear matter and is captured by even the simplest liquid drop models. However, when inspected in detail, nuclear structure plays a significant role in modulating $r$-process isotope production \cite{mumpower}. 

The most prominent structure effects are the major shell closures at $N$~=~50,~82,~and~126, which cause bottlenecks in the $r$-process flow and enhanced abundance of elements produced at these locations \cite{langanke}. Beyond that, smaller effects, like shell-driven areas of large deformation, shape coexistence, nuclear isomers, and anomalously slow $\beta$ decays (caused by large spin differences, or poor overlap of parent and daughter wave functions) result in more modest modulations in the final $r$-process stable-isotope production. The exact locus of the $r$-process is still not accurately known, and most nuclei on the expected path are yet to be produced and measured. Experimental study of these nuclei is a major goal of new, `next-generation' radioactive-beam facilities currently under construction. Many important cases are refractory elements, whose production is suppressed with current Isotope Separation On-Line (ISOL) techniques. However, a growing number of recent results have yielded a wealth of nuclear-structure information and considerable progress is being made in pushing into this neutron-rich region with existing infrastructure, motivated by both astrophysical and nuclear-structure reasons. 

This specific research is aimed at clarifying the mass and spin of highly deformed \nuc{106}{Nb}, and at seeking a long-lived, low-lying $\beta$-decaying isomer, similar to those found in \nuc{100,102,104}{Nb}. Such isomers are ubiquitous in odd-odd nuclei in the region; a consequence of near-degenerate structures of pure $pf$-shell, or $g$-shell, parentage. The structure of the progeny, \nuc{106}{Mo}, has been well investigated through prompt-fission-fragment $\gamma$-ray spectroscopy, but our $\beta$-decay study populated a wealth of new low-spin levels and offers access to particle-hole states not seen in prompt fission. During the preparation of this manuscript, a similar $\beta$-decay study performed at the RIKEN RI Beam Factory was published \cite{ha}. The results presented below are in broad agreement with the findings of the RIKEN work, although some details differ, both in the data and in their interpretation.


\section{EXPERIMENT DETAILS}\label{sec:experiment}

This work was performed at the CAlifornium Rare Isotope Breeder Upgrade (CARIBU) facility at Argonne National Laboratory. Here, neutron-rich radioactive nuclei produced in the spontaneous fission of \nuc{252}{Cf} are extracted and thermalised in the CARIBU gas catcher. The species of interest is mass-selected by an isobar separator, bunched, and delivered to the required experimental area. Details relevant to the reported experiments are provided below. For a more detailed description of the CARIBU facility, we refer the reader to existing literature, for example Ref.~\cite{savard}. Here, we report on the first dedicated inspection of the ground-state and decay properties of \nuc{106}{Nb} via complementary nuclear mass measurements and $\beta$-delayed $\gamma$-ray spectroscopy. 

\subsection{CANADIAN PENNING TRAP}\label{subsec:cpt}

A mass measurement was performed using the Canadian Penning Trap (CPT) \cite{vanschelt} to confirm the accuracy of the reported \nuc{106}{Nb} ground-state mass \cite{mwang}. At CARIBU, \nuc{106}{Nb} ions were extracted from the gas catcher in a 2$^+$ charge state, and a bunched beam was produced at a repetition rate of 10~Hz. To remove unwanted contaminant ions from the beam, the new Multi-Reflection Time-Of-Flight (MR-TOF) mass separator \cite{hirsh} was employed. Ion bunches were captured in the MR-TOF and allowed to isochronously cycle between the two ion mirrors for a duration of 10~ms, wherein a mass resolving power of $R = m/\Delta m > 50,000$ was achieved. A Bradbury-Nielsen Gate \cite{bradbury} at the MR-TOF exit was used to selectively transfer \nuc{106}Nb$^{2+}$ ions to the low-energy experimental area, while suppressing other $A$~=~106 isobars by several orders of magnitude. 

The resulting ion bunches were collected in a cryogenic linear RFQ trap, where they were cooled and re-bunched for injection into the Penning trap. The mass measurement was conducted using the Phase-Imaging Ion-Cyclotron-Resonance (PI-ICR) technique \cite{eliseev}. In this method, a position-sensitive micro-channel plate is used to infer the phase of the orbital motion of trapped ions at some given time. The cyclotron frequency ($\nu_c$) is determined by measuring the change in phase during a period of excitation-free accumulation ($t_{acc}$). After time $t_{acc}$ in the Penning trap, the ions are ejected and the position of the ions at the detector plane is measured. Ions acquire a mass-dependent phase during the accumulation time and form clusters (or \textit{spots}) at some radius from the projected trap centre. The angle between these spots and a mass-independent reference spot is measured ($\phi_c$) and the cyclotron frequency is given by: \\[-0.7cm]

\begin{equation}
\nu_c = \frac{\phi_c + 2\pi N}{2\pi t_{acc}},
\end{equation}

\noindent
where $N$ is the integer number of revolutions during the time $t_{acc}$. The technique provides high sensitivity and resolution, and is therefore also well-suited to search for low-lying or weakly produced isomers. A 1-s accumulation time results in a mass resolution of $R\approx 1.5\times10^7$. Details of the implementation of this measurement technique at the CPT are introduced in Refs.~\cite{orford1, orford2}. 

\subsection{X-ARRAY AND SATURN DECAY-SPECTROSCOPY STATION}\label{subsec:xarray}

The $\beta$-decay properties of \nuc{106}{Nb} were investigated using the X-Array and SATURN decay-spectroscopy station \cite{mitchell1}. The decay-spectroscopy station consists of up to five high-efficiency High-Purity Germanium (HPGe) clover-style $\gamma$-ray detectors, and a plastic scintillator offering almost complete solid-angle coverage. The system has been demonstrated to be a powerful spectroscopy device with low-intensity, radioactive-ion beams \cite{mitchell2}. A low-energy beam of mass-separated \nuc{106}{Nb} ions, bunched and delivered at 100-ms intervals, was deposited on a movable aluminized-mylar tape located in the geometric centre of the array at a rate of 100-200~ions/second. The X-Array configuration described in Ref.~\cite{mitchell1} was modified slightly for this experiment. The clover detector located on the left-hand-side of the X-Array, as observed by the oncoming beam particles, was removed and replaced with five unshielded LaBr$_3$ scintillators. The purpose here was to test the capacity of the modified X-Array to measure excited-state lifetimes. Unfortunately, due to the high level of room-background, no useful information was extracted from the LaBr$_3$ detector data, and so these are not discussed any further here.  

Despite the MR-TOF described above not being available at the time, the beam delivered for this experiment consisted primarily of mass-selected \nuc{106}{Nb} ions. Small contributions from neighbouring isobars, \nuc{106}{Zr} and \nuc{106}{Mo}, may be expected due to the small mass differences and the maximum achievable mass resolution of the isobar separator at the time of this experiment. However, the presence of \nuc{106}{Zr} is effectively suppressed due to the relative proportion of its spontaneous fission branch and the low intensity of the radioactive-ion beam. There are no known $\gamma$ rays associated with \nuc{106}{Zr}~$\rightarrow$~\nuc{106}{Nb} $\beta$ decay for identification. Six \nuc{106}{Nb} $\gamma$-ray transitions with relative intensities $>$~10$\%$ are known from prompt-fission spectroscopy \cite{defrenne}; these were undetectable in both the $\gamma$-ray singles and coincidence data. Any beam contamination leading directly to \nuc{106}{Mo}~$\rightarrow$~\nuc{106}{Tc} decay would be suppressed along with the other long-lived isobaric contamination by the repeating beam cycle, described below, that was applied throughout the experiment.

Data were collected in two modes of repeating tape-movement cycles: one lasted for 14.0~s; the other for 7.5~s. The growth-and-decay collection cycle of alternating `beam on' and `beam off' periods was achieved by switching an electrostatic beam deflector with the SATURN logic control system. The implantation tape was moved at the end of each cycle to suppress accumulation of activity from long-lived decay products at the collection site. The longer cycle was used to measure the \nuc{106}{Nb} decay half-life; this technique was successfully demonstrated in the earlier work of Ref.~\cite{siegl}. The shorter cycle was adopted to maximise the collection rate for \nuc{106}{Nb} decay. While isobaric contamination of the $\gamma$-ray spectra was suppressed by the moving tape cycle, the relatively short half-lives involved meant that some level of contamination was unavoidable. Over time, activity build-up on the tape led to contribution of isobaric $\beta$ decay from \nuc{106}{Mo}~$\rightarrow$~\nuc{106}{Tc} (T$_{1/2}$~=~8.73(12)~s) and \nuc{106}{Tc}~$\rightarrow$~\nuc{106}{Ru} (T$_{1/2}$~=~35.6(6)~s). Since the half-life of \nuc{106}{Ru} is T$_{1/2}$~=~371.8(18)~days \cite{defrenne}, this was effectively the end of the decay chain over the days-long timescale of this experiment. The photopeak of the most-intense $\gamma$-ray transition observed in \nuc{106}{Mo} is five-to-six times larger than the corresponding transitions in \nuc{106}{Tc} and \nuc{106}{Ru}. In many cases, it was possible to confirm assignments of new $\gamma$~rays to the appropriate isobar by measuring the associated $\beta$-decay half-life. 

Standard $\gamma$-ray sources of \nuc{243}{Am}, \nuc{56}{Co}, \nuc{152}{Eu}, and \nuc{182}{Ta} were used to calibrate the detection efficiency of the X-Array up to $\approx$3.5~MeV. Well-known, room-background $\gamma$ rays were also used to obtain an energy calibration exceeding the range of interest for this experiment (which was E$_{\gamma}$~$\approx$3~MeV). In particular, high-energy $\gamma$ rays produced from ($n$,$\gamma$) reactions, a consequence of the high neutron flux emitted from the CARIBU \nuc{252}{Cf} source, were used to confirm the appropriate use of a linear calibration. Photopeaks of these $\gamma$ rays appear in the $\gamma$-ray singles data, but are removed by applying a $\beta$- or $\gamma$-coincidence condition in offline data sorting. Systematic uncertainty of the energy calibration was found to be $\lesssim$~0.1~keV. The uncertainties of measured $\gamma$-ray energies quoted in this work include the systematic uncertainty, as well as the statistical uncertainty associated with the fitting routines of the $gf3$ software package \cite{radware}. The measured energy resolution of the X-Array in this work was 2.5~keV at 1000~keV, 3.7~keV at 2000~keV and 4.2~keV at 3000~keV.

Data were collected using a digital acquisition system (DAQ) that applied a free-running trigger. Signals from the individual clover crystals and tape-cycle reset trigger were input directly in the DAQ. The outputs of three Hammamatsu PMTs associated with the BC-408 plastic-scintillator detector in SATURN were coupled together and amplified before being delivered to the DAQ. Data were sorted offline into a combination of singles spectra and coincidence matrices that were used in the subsequent analyses discussed below. 


\section{GROUND-STATE PROPERTIES OF \nuc{\bf{106}}{Nb}}\label{sec:gsp}

\subsection{GROUND-STATE MASS}\label{subsec:mass}

\begin{figure}[t]
\includegraphics[width=8.5cm]{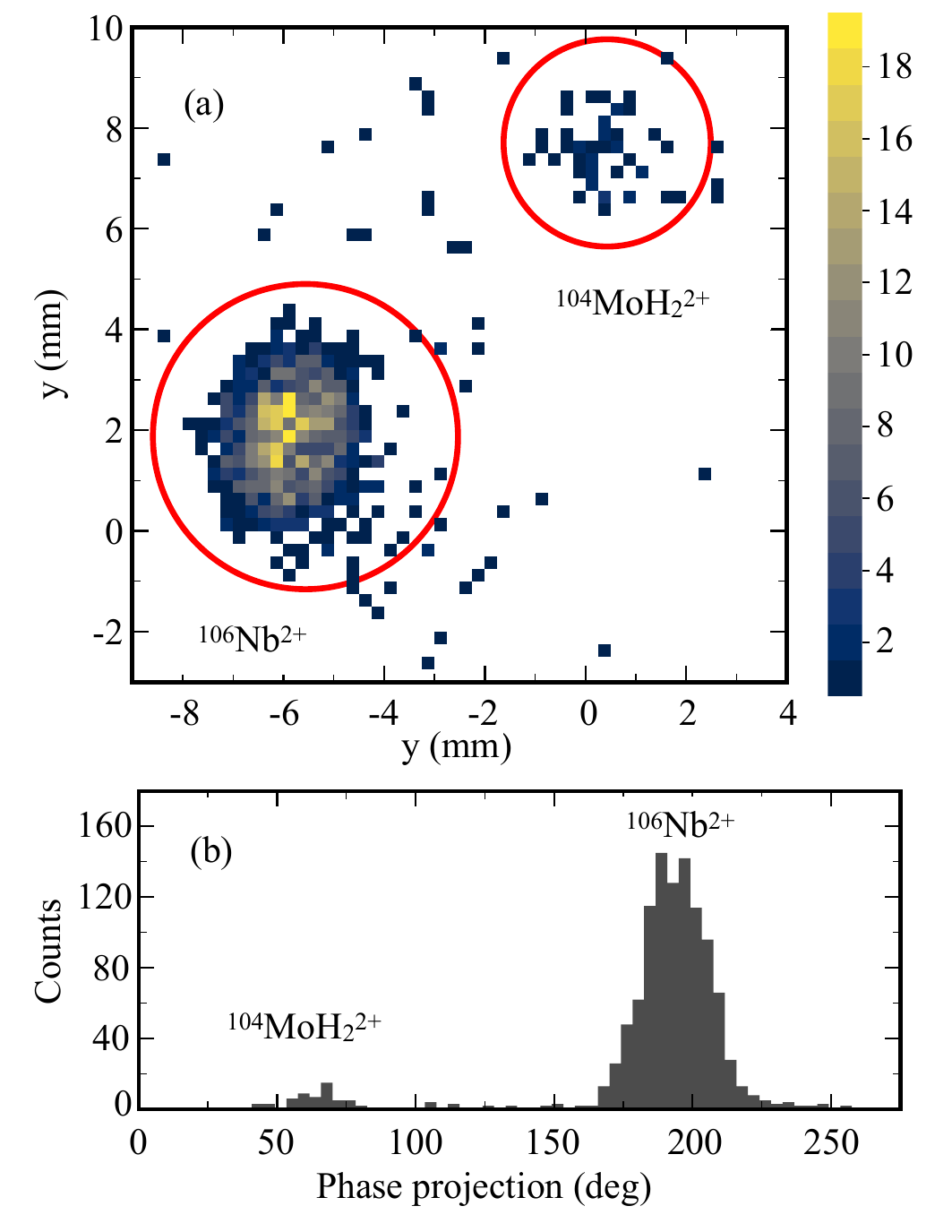}
\caption{\label{fig1} Example CPT spectra acquired using the PI-ICR technique with $t_{acc}$~=~190~ms. (a) Ions acquire a mass-dependent phase, forming characteristic `\textit{spots}', during the collection time in the trap; the \nuc{106}{Nb}$^{2+}$ and molecular \nuc{104}{Mo}H$_2$$^{2+}$ are identified. (b) Corresponding phase projection of \nuc{106}{Nb}$^{2+}$ and the \nuc{104}{Mo}H$_2$$^{2+}$ contaminant. }
\end{figure}

\begin{figure}[t]
\includegraphics[trim=0cm 6cm 0cm 0cm, clip=true, width=8.5cm]{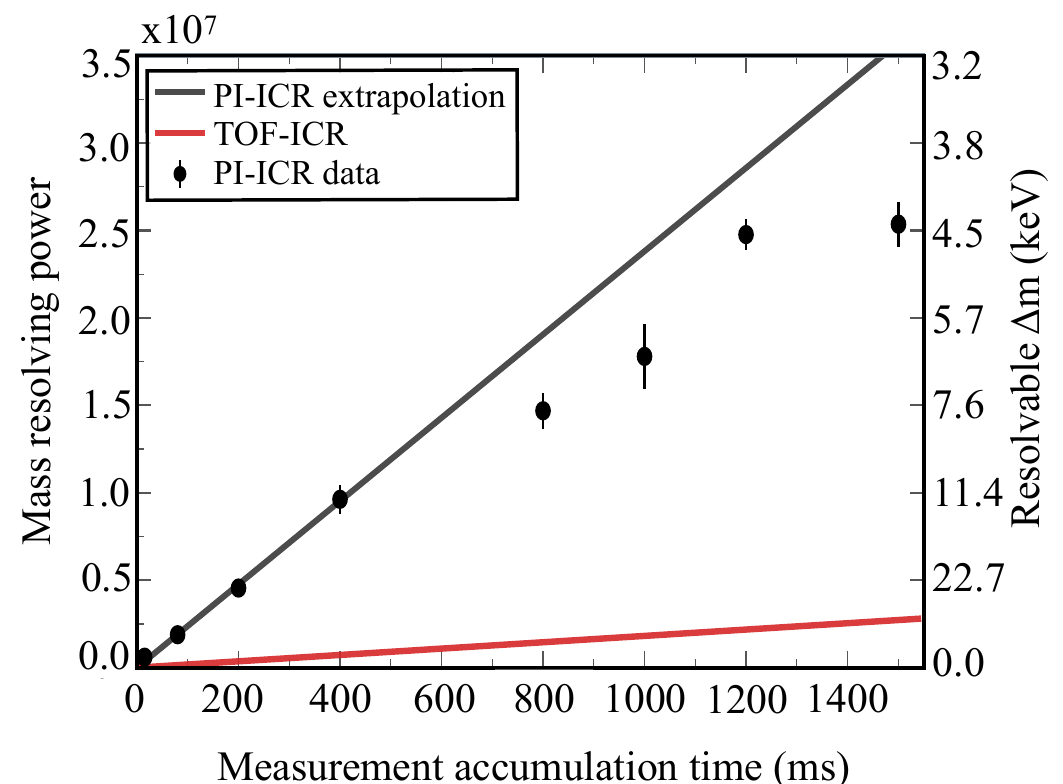}
\caption{\label{fig2} Mass resolving power and resolvable mass differences with the PI-ICR technique (black line) as a function of accumulation time, $t_{acc}$. For comparison, the achievable resolving power with the Time-of-Flight Ion-Cyclotron-Resonance (TOF-ICR) technique (red line) is also shown.}
\end{figure}

The CPT system was calibrated by measuring the cyclotron frequency of \nuc{52}Cr$^+$, which is readily available at CARIBU and has a precisely known mass \cite{mwang}. To reduce systematic uncertainties, the calibration was performed under the same experimental conditions as the \nuc{106}{Nb} mass measurement, using the same accumulation times. A single contaminant species, \nuc{104}MoH$_2^{2+}$, was identified in the \nuc{106}Nb$^{2+}$ beam with an intensity roughly 20 times weaker than the collected \nuc{106}Nb ions. Accumulation times were chosen such that the contaminant molecule and \nuc{106}Nb were completely resolved in the measured spectra. 

Measurement of the \nuc{106}{Nb} cyclotron frequency was achieved from several phase-accumulation times near 190~ms. An example phase-measurement spectrum is provided in Fig.~\ref{fig1}. With the PI-ICR technique, an increase in the accumulation time results in a corresponding increase in mass resolving power of the measurement; this is presented in Fig.~\ref{fig2}. As $t_{acc}$ increases, the spot size FWHM also increases, which results in the drop-off from the extrapolation line. If a long-lived, excited state were to occur in \nuc{106}Nb within approximately 30~keV of the ground state, it could be partially obscured by the spot for $t_{acc}$~$\approx$~190-ms accumulation. In this work, the accumulation time was scanned between approximately 15~ms~$\leq~t_{acc}~\leq$~1500~ms, with several intermediate steps, to search for any unknown, long-lived (T$_{1/2}$~$\geq$~10~ms) excited states in \nuc{106}{Nb}. As the corresponding mass resolving power surpasses the physical mass difference between the ground state and any possible isomer, the two would separate into resolved spots. The evolution of the spot FWHM with accumulation time was within the tolerance that is expected due to Penning trap voltage instabilities, resulting in an exclusion limit of $\leq$~5~keV on the excitation energy of any potential long-lived isomer. From the measured cyclotron frequency, the ground-state mass of \nuc{106}{Nb} was found to be $-$66202.0(13)~keV, which is in agreement with the value of $-$66203(4)~keV from Ref.~\cite{hager} which was adopted in the 2016 Atomic Mass Evaluation \cite{mwang}. In the previous work, the masses of several Nb isotopes, including \nuc{106}Nb, were measured with the JYFLTRAP double Penning trap \cite{hager}. In that experiment, the expected isomer in \nuc{104}Nb was not observed, and there is no mention of a search for an isomer in \nuc{106}Nb.

\subsection{\boldmath{$\beta$}-DECAY HALF-LIFE}\label{subsec:thalf}

\begin{figure}[t]
\includegraphics[trim=0cm 0cm 0cm 0cm, clip=true, width=8.5cm]{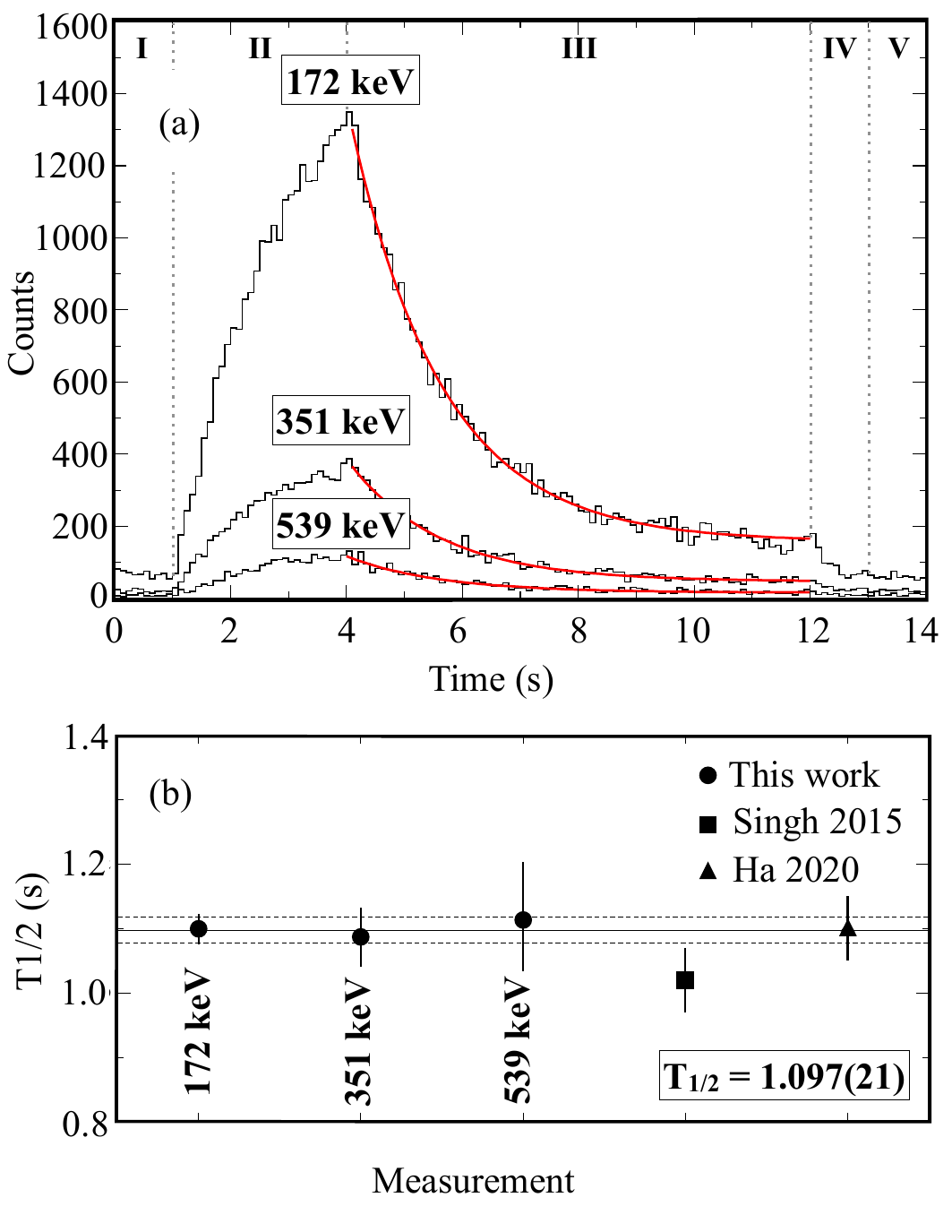}
\caption{\label{fig3} (a) Illustration of the 14-s beam cycle used in the experiment. The data are gated on the 172-keV (2$^+_1$$\rightarrow$~0$^+_1$), 351-keV (4$^+_1$$\rightarrow$~2$^+_1$), and 539-keV (2$^+_2$$\rightarrow$~2$^+_1$) transitions in \nuc{106}{Mo}. Different stages of the time cycle are indicated at the top of the figure: (I) Room background; (II) Beam-on collection; (III) Beam-off collection; (IV) Mylar tape movement; and (V) Room background. Exponential functions fit to the `beam-off' period are shown for each individual $\gamma$-ray transition. (b) The measured half-lives are provided along with the updated evaluation of Ref.~(Singh 2015:~\cite{singh}) and recent measurement of Ref.~(Ha 2020:~\cite{ha}). The weighted mean (solid line) $\pm$ 1$\sigma$ (dashed lines) of the three individual measurements from this work gives a value of T$_{1/2}$~=~1.097(21)~s, which is consistent with the work of Ha~$et~al$. \cite{ha} (1.10(5)~s) but is $\approx$8$\%$ larger than the adopted value (1.02(5)~s).}
\end{figure}

The most-recent NNDC evaluation of \nuc{106}{Nb} \cite{singh} reports a $\beta$-decay half-life of T$_{1/2})$~=~1.02(5)~s. This is the value reported in Ref.~\cite{shizuma} from decay curves for the 172- and 351-keV transition; other values ranging from 0.90(2)~s to 1.240(21)~s from the references stated therein are excluded by the evaluator. Application of a repeating on and off data-collection cycle, in phase with beam delivery to the spectroscopy station, allowed the $\beta$-decay half-life of \nuc{106}{Nb} to be measured in this work with greater precision. Data were sorted into a two-dimensional matrix of HPGe $\gamma$-ray time relative to the beginning of the data-collection cycle versus the measured energy of that $\gamma$ ray. Exponential decay curves were obtained by applying a cut on individual $\gamma$-ray energies and projecting the data onto the timing axis. The decay half-life was obtained by fitting an exponential function with a constant background to the beam-off portion of the cycle (indicated in Fig.~\ref{fig3}). This process is presented for three $\gamma$-ray transitions that depopulate low-lying excited states in \nuc{106}{Mo}, namely the 172-keV (2$^+_1$$\rightarrow$~0$^+_1$), 351-keV (4$^+_1$$\rightarrow$~2$^+_1$), and 539-keV (2$^+_2$$\rightarrow$~2$^+_1$) transitions. A weighted mean of these values suggests that the $\beta$-decay half-life of \nuc{106}{Nb} is T$_{1/2}$~=~1.097(21)~s. The larger uncertainties of the data points for E$_{\gamma}$=~351, 539~keV are reflective of lower statistics. This result is consistent with recent measurement of Ha~$et~al$  \cite{ha}, which has a larger uncertainty (T$_{1/2}$~=~1.10(5)~s). The improved precision points to a discrepancy of $\approx$8$\%$ with the current adopted value of 1.02(5)~s \cite{singh}. 

\subsection{APPARENT {\boldmath $\beta$}-DECAY FEEDING}\label{subsec:bfeeding}

\begin{figure}[t]
\includegraphics[trim=0cm 5.5cm 0cm 0cm, clip=true, width=8.5cm]{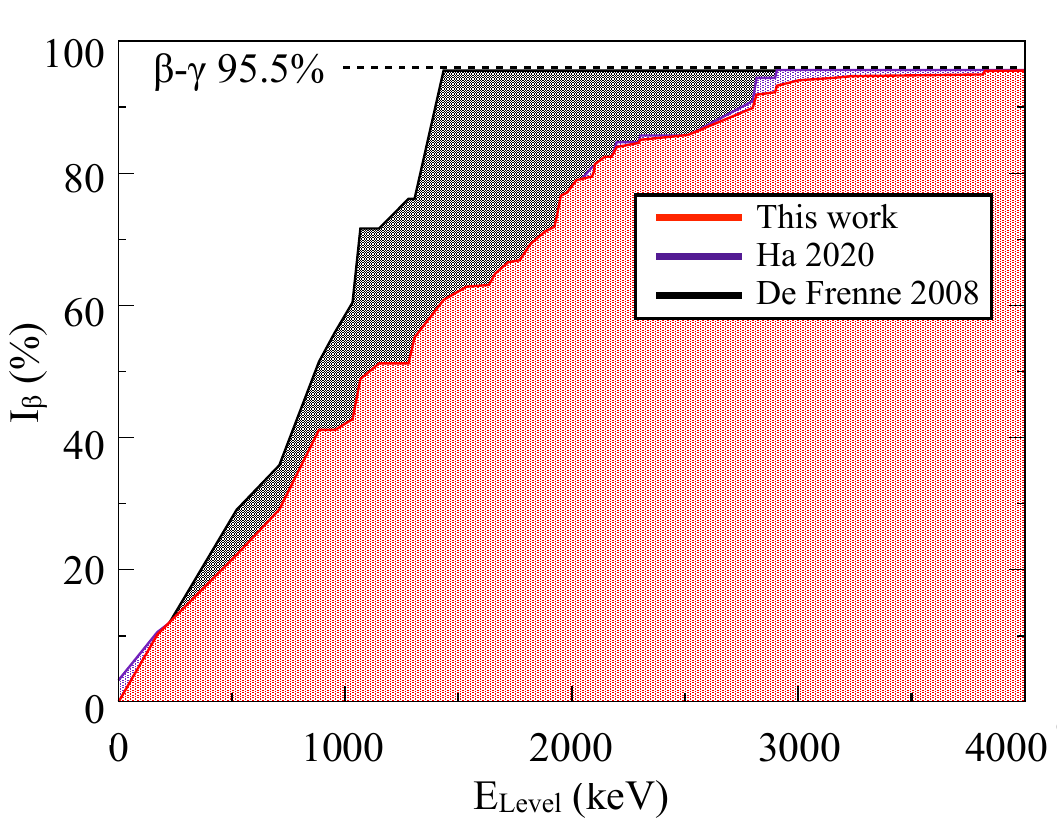}   
\caption{\label{fig4} Accumulation of the apparent $\beta$-feeding strength of \nuc{106}{Nb} as a function of excitation energy of the decay progeny, \nuc{106}{Mo}, from this work (red), Ha $et~al$.~(Ha 2020:~\cite{ha}) (purple) and derived from $\gamma$-ray intensities given in the most-recent data evaluation~(De Frenne 2008:~\cite{defrenne}) (black). A $\beta$-delayed neutron branch of 4.5(3)$\%$ for \nuc{106}{Nb} is assumed \cite{defrenne}.}
\end{figure}

Apparent $\beta$-decay feeding intensities have been obtained through a balance of the measured $\gamma$-ray intensities that feed and depopulate each level; the expanded level scheme is discussed in detail below. A $\beta$-delayed neutron-emission branch of 4.5(3)$\%$ for \nuc{106}{Nb} is reported in the literature (see Refs.~\cite{gomez-hornillos,pereira}, for example). Several \nuc{105}{Mo} $\gamma$ rays \cite{lalkovski} were identified in the coincidence data by setting gates at energies corresponding to transitions in this nucleus. For example, the strongest transition that depopulates the first excited state at 95~keV is of mixed $M$1+$E$2 character, with mixing ratio $\delta$~=~-0.24(4) and total internal conversion coefficient $\alpha$~=~0.355(22) \cite{lalkovski}. A coincidence gate on this $\gamma$ ray revealed the two strongest transitions (when fed from \nuc{105}{Nb} $\beta$ decay) at 138~keV and 254~keV. For reference, $I^{105}_{\gamma}$(254)~$\approx$~1$\%$[$I^{106}_{\gamma}$(172)]. No $\gamma$ rays from \nuc{105}{Mo}~$\rightarrow$~\nuc{105}{Tc} $\beta$ decay were observed. 

The total apparent $\beta$ feeding to excited states in \nuc{106}{Mo} was normalized to account for the adopted $\beta$-delayed neutron branch; accumulation as a function of level excitation energy is presented in Fig.~\ref{fig4} for this work, along with that of Ref.~\cite{ha} and Refs.~\cite{defrenne, shizuma}. This highlights the all-too-common deficiencies of limited historical data available in the literature, particularly concerning the decay properties of neutron-rich isotopes in this region. The adopted levels \cite{defrenne, shizuma} suggest that the average energy released from relaxation of the decay product, weighted by the quoted $\beta$-feeding intensities, is $\approx$950~keV. In the proposed decay scheme of Ref.~\cite{ha}, this value increases by approximately 30$\%$ to $\approx$1300~keV, which is similar to the feeding distribution observed in this work. 

Further still, the large $\beta$-decay $Q$ value of 9.931(10)~MeV and lack of excited states observed above 4~MeV implies that the Pandemonium effect \cite{hardy} may be strong in this nucleus. Direct feeding of high-energy states embedded in a region of high level density would result in a cascade of low-energy, low-intensity $\gamma$ rays that are below the threshold of sensitivity for this measurement. As a result, the individual apparent $\beta$-feeding intensities are quoted as upper limits in Table~\ref{summary}. Using the measured decay half-life, $\beta$-feeding intensities and adopted $Q$ value, log-$ft$ values have been calculated using the NNDC LOGFT program \cite{logft}. The range of extracted log-$ft$ values, $\approx 6.0-7.0$, suggests that the observed excited states in \nuc{106}{Mo} are most likely populated via a series of allowed or first-forbidden $\beta$ decays.

Since the adopted ground-state spin-parity assignment of \nuc{106}{Nb} is \jp{1}{-}{} \cite{defrenne}, the $\beta$-feeding pattern should be dominated by allowed Gamow-Teller and Fermi decays to \jp{0,1,2}{-}{} states in \nuc{106}{Mo}, which must lie above the pairing gap in the even-even decay product. One would expect these states to be connected to the lowest-lying levels via electric dipole decays; however, this is not the case. Also, we do not report any excited 0$^+$ states in this work, while only a modest fraction of the observed $\beta$ feeding proceeds to known 2$^+$ levels. In fact, it was surprising to find that at least half of the observed $\beta$ feeding was to known states of spin $J={3-5}$. This distribution of apparent $\beta$-feeding strength appears to rule out a \jp{1}{-}{} assignment for the \nuc{106}{Nb} ground state, and is discussed in further detail below. 

\begin{figure*}[ht!]
\includegraphics[trim=0cm 6.2cm 0cm 0cm, clip=true, width=18cm]{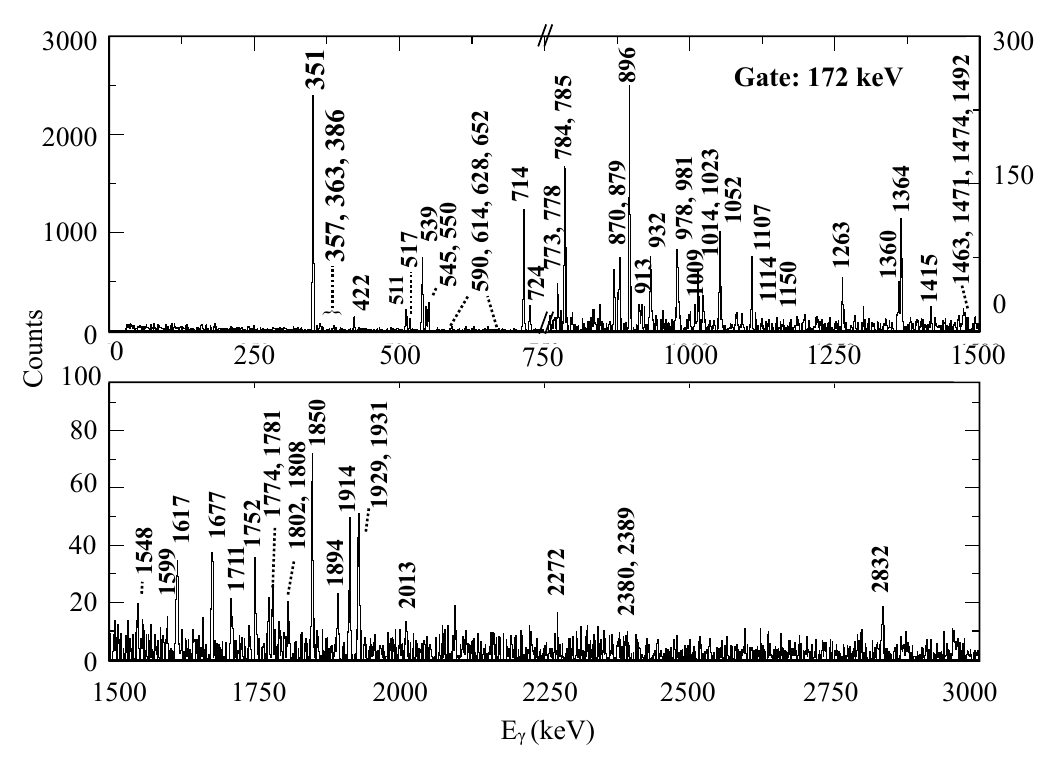}
\caption{\label{fig5} Background-subtracted, $\beta$-gated, $\gamma$-$\gamma$-coincidence matrix, gated on the well-known 172-keV (2$^+_1$$\rightarrow$~0$^+_1$) transition in \nuc{106}{Mo}, from (top) 0~keV to 1500~keV, and (bottom) 1500~keV to 3000~keV. The $\gamma$ rays from transitions in \nuc{106}{Mo} are labelled with their energies. Note the change of y-axis scale at 750~keV in the top panel.}
\end{figure*}

\section{OBSERVED {\boldmath $\gamma$} DECAY of \nuc{\bf{106}}{Mo}}\label{sec:gammas}

\begin{figure*}[t!]
\includegraphics[trim=0cm 6.2cm 0cm 0cm, clip=true, width=18cm]{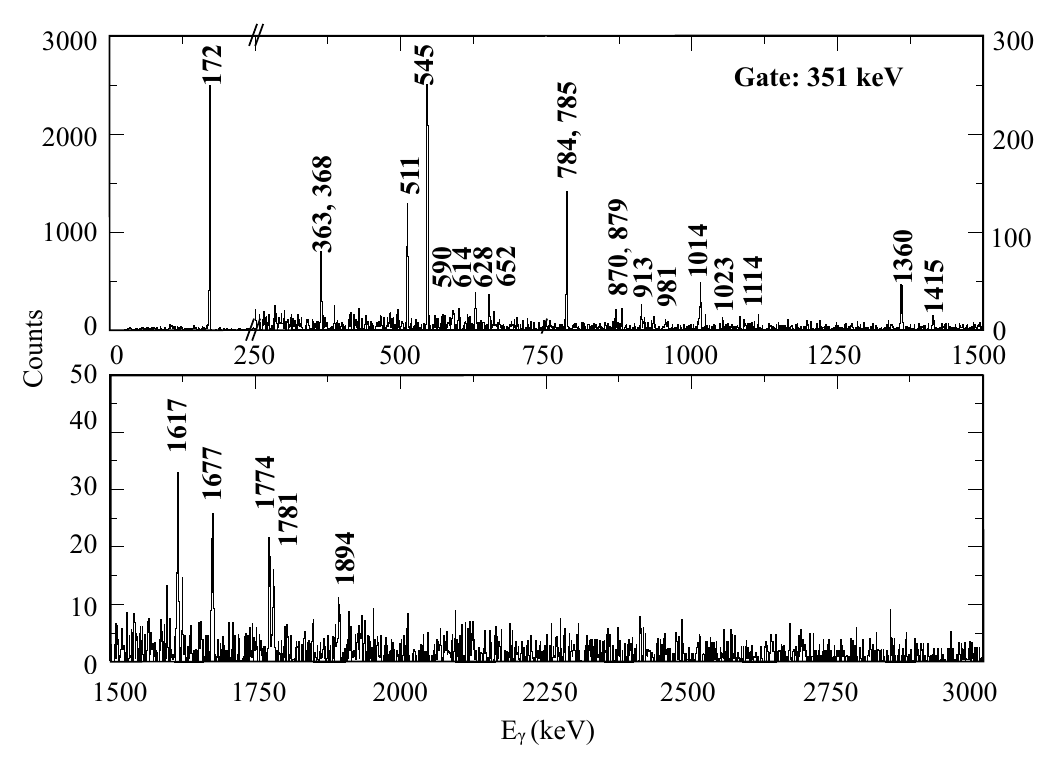}
\caption{\label{fig6} Background-subtracted, $\beta$-gated, $\gamma$-$\gamma$-coincidence matrix, gated on the well-known 351-keV (4$^+_1$$\rightarrow$~2$^+_1$) transition in \nuc{106}{Mo}, from (top) 0~keV to 1500~keV, and (bottom) 1500~keV to 3000~keV. The $\gamma$ rays from transitions in \nuc{106}{Mo} are labelled with their energies. Note the change of y-axis scale at 250~keV in the top panel.}
\end{figure*}

\begin{figure}[t!]
\includegraphics[trim=0cm 0cm 0cm 0cm, clip=true, width=8.5cm]{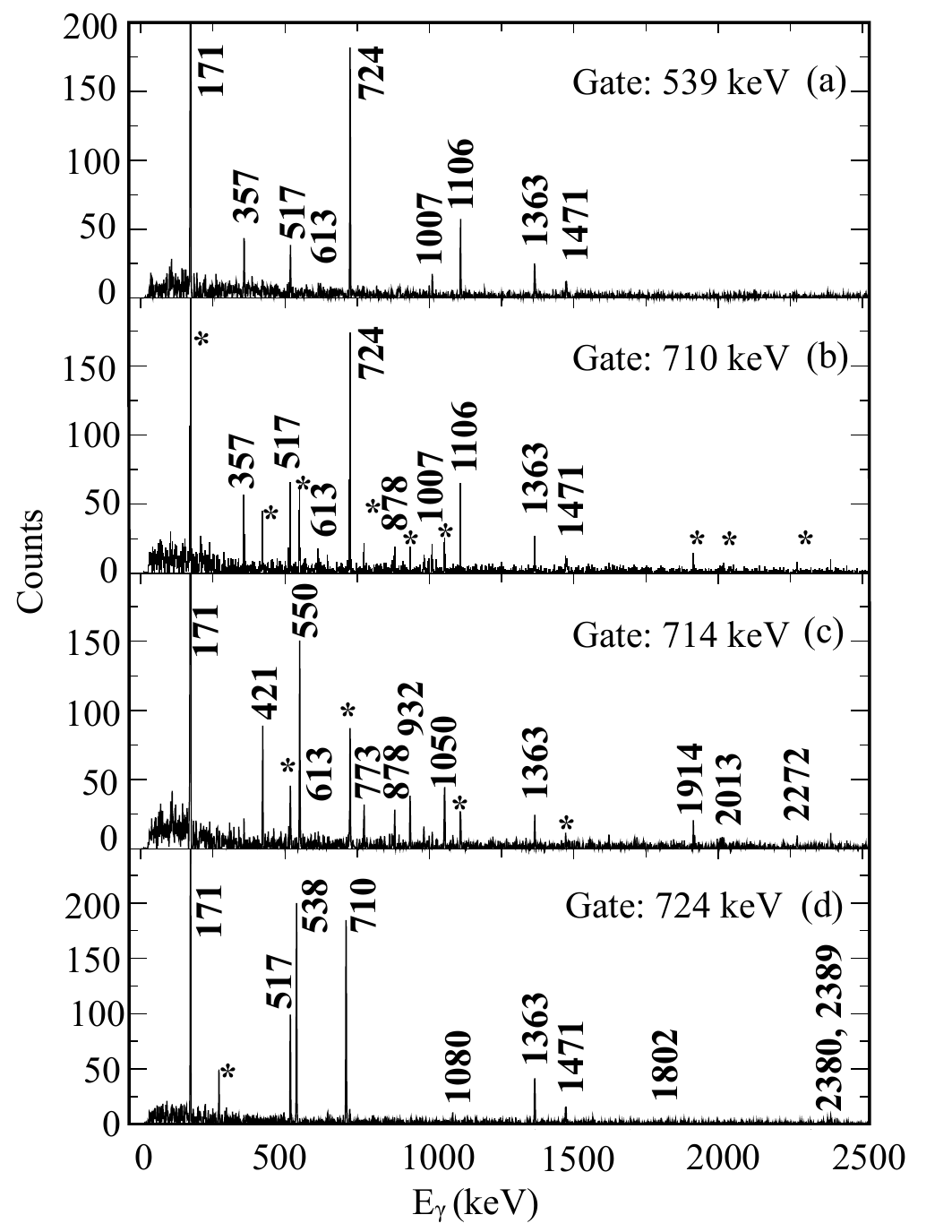}
\caption{\label{fig7} Background-subtracted, $\beta$-gated, $\gamma$-$\gamma$-coincidence matrix, gated on the established (a) 539-keV (2$^+_2$$\rightarrow$~2$^+_1$), (b) 710-keV (2$^+_2$$\rightarrow$~0$^+_1$), (c) 714-keV (3$^+_1$$\rightarrow$~2$^+_1$), and (d) 724-keV (4$^+_2$$\rightarrow$~2$^+_2$) transitions in \nuc{106}{Mo}, from 0~keV to 2500~keV. The $\gamma$ rays from transitions in \nuc{106}{Mo} are labelled with their energies. A * indicates contamination from the energy gates overlapping nearby $\gamma$ rays.}
\end{figure}

\begin{figure}[t!]
\includegraphics[trim=0cm 0cm 0cm 0cm, clip=true, width=8.5cm]{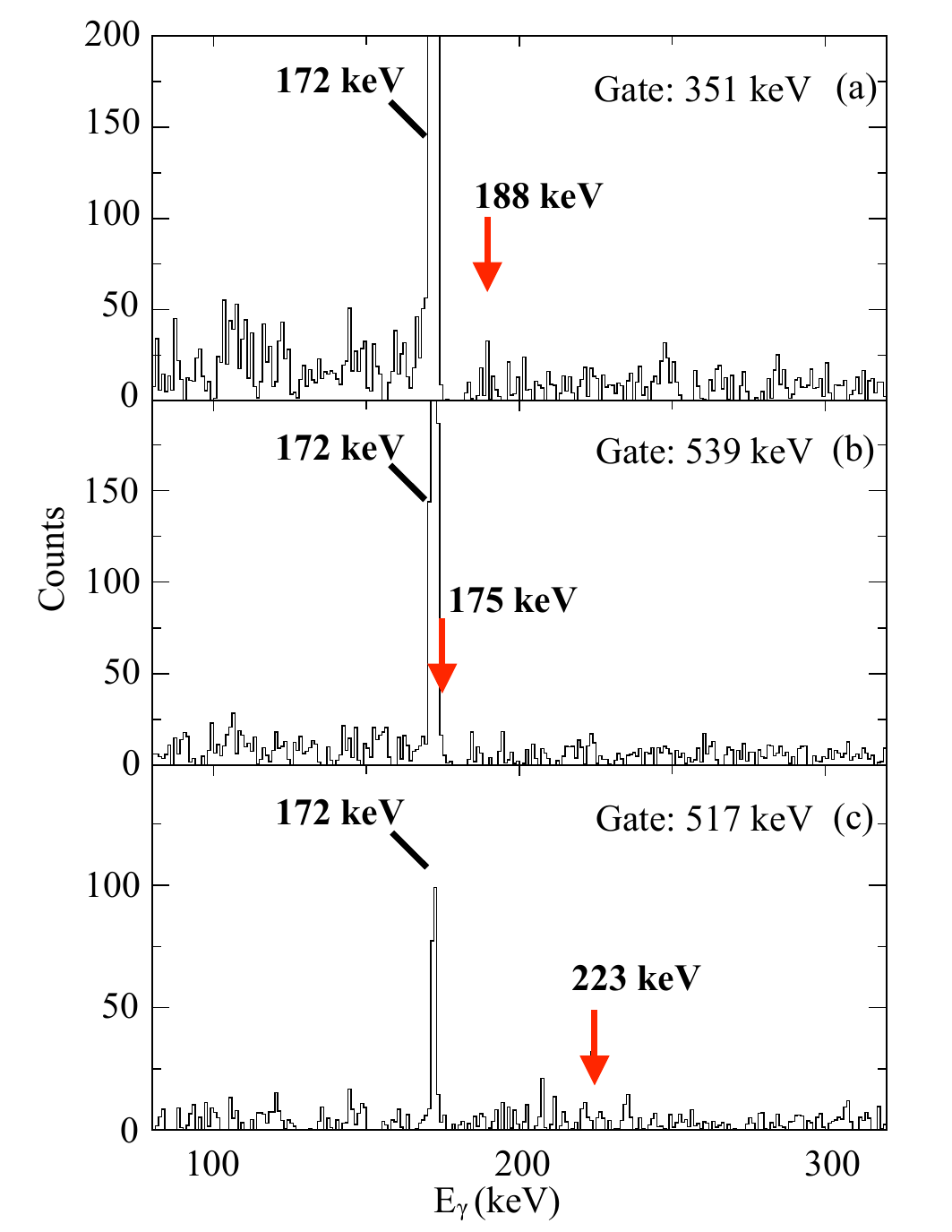}
\caption{\label{fig8} Background-subtracted projection of the $\beta$-gated, $\gamma$-$\gamma$-coincidence matrix, gated on the (a) 351-keV (4$^+$$\rightarrow$~2$^+$), (b) 539-keV (2$^+$$\rightarrow$~2$^+$), and (c) 517-keV (J$^{\pi}$~$\rightarrow$~5$^+_{(1)}$) transitions in \nuc{106}{Mo}, from 100~keV to 300~keV. Expected locations of the unobserved $\gamma$ rays from Ref.~\cite{ha} are indicated by the red arrows and discussed in the text.}
\end{figure}

Observed $\gamma$ rays were assigned to \nuc{106}{Mo} through inspection of $\gamma-\gamma$ coincidence relationships and $\beta$-decay half-life measurements. Placement of $\gamma$ rays in the \nuc{106}{Mo} decay scheme was achieved through gating on known transitions that strongly depopulate low-lying excited states. Examples of background-subtracted projections of the $\gamma$-$\gamma$ coincidence matrix used in this work, gated on transitions that depopulate the established 172-keV (\jp{2}{+}{1}), 351-keV (\jp{4}{+}{1}), 710-keV (\jp{2}{+}{2}), 885-keV (\jp{3}{+}{1}), and 1435-keV (\jp{4}{+}{2}) levels are presented in Figs.~\ref{fig5}, \ref{fig6}, and \ref{fig7}, respectively. Where possible, the locations of excited states, and transitions that connect them, were confirmed by applying $\gamma$-ray coincidence gates to transitions lying higher in the level scheme. The same techniques were applied to confirm the identification of isobaric contamination in the data. 

Most relative $\gamma$-ray intensities, $I_{\gamma}$, were determined by gating on a transition that depopulates the level to which the $\gamma$ ray under inspection is directly feeding. Photopeak yields measured in the coincidence spectra were corrected for their $\gamma$-ray detection efficiency, the gating transition detection efficiency and branching-ratio fraction, and, in the case of the 172-keV gate, internal conversion. A theoretical conversion coefficient of 0.171(2) was calculated for this transition using the BRICC code \cite{kibedi}, assuming that it is a pure $E$2 transition. Internal conversion is expected to have a small, or negligible contribution for almost all of the other transitions with higher energies; for example, the total conversion coefficient is $\approx$1$\%$ for the 351-keV (4$^+_1$$\rightarrow$2$^+_1$) transition. Different approaches were taken for the three transitions that feed directly to the ground state: $I_{\gamma}$(172) was determined from the $\beta$-gated $\gamma$-ray singles data; $I_{\gamma}$(710, 1150) were found by gating on transitions that feed into these excited states. The measured branching ratios of these two $\gamma$-ray transitions were consistent with the corresponding $I_{\gamma}$ values measured from $\beta$-gated singles data. The $I_{\gamma}$(172) values from this work are reported in Table~\ref{summary}, with the 172-keV transition normalised to 100 units.

\subsection{EXCITED STATES OF \nuc{\bf{106}}{Mo}}\label{subsec:mo106}

The work of Shizuma $et~al.$ in 1983 \cite{shizuma} was the first to exploit $\beta$ decay of \nuc{106}{Nb} as a means to investigate the level structure of \nuc{106}{Mo}. For almost 40 years, this remained the only $\beta$-delayed $\gamma$-ray spectroscopy of \nuc{106}{Mo} reported in the literature. Structurally, much of what is known on \nuc{106}{Mo} has come through high-fold, $\gamma$-ray spectroscopy of prompt fission fragments with preferential population of high-spin states and extended rotational bands \cite{hua,jones,ruiqing}. At the time of writing, Ha~$et~al.$ \cite{ha} examined the role of triaxiality in \nuc{106-110}{Mo} via the $\beta$-decay of \nuc{106-110}{Nb}, extending the known level schemes of each isotope.

\begin{figure*}[t]
\includegraphics[trim=0cm 1.5cm 0cm 0cm, clip=true, width=17cm]{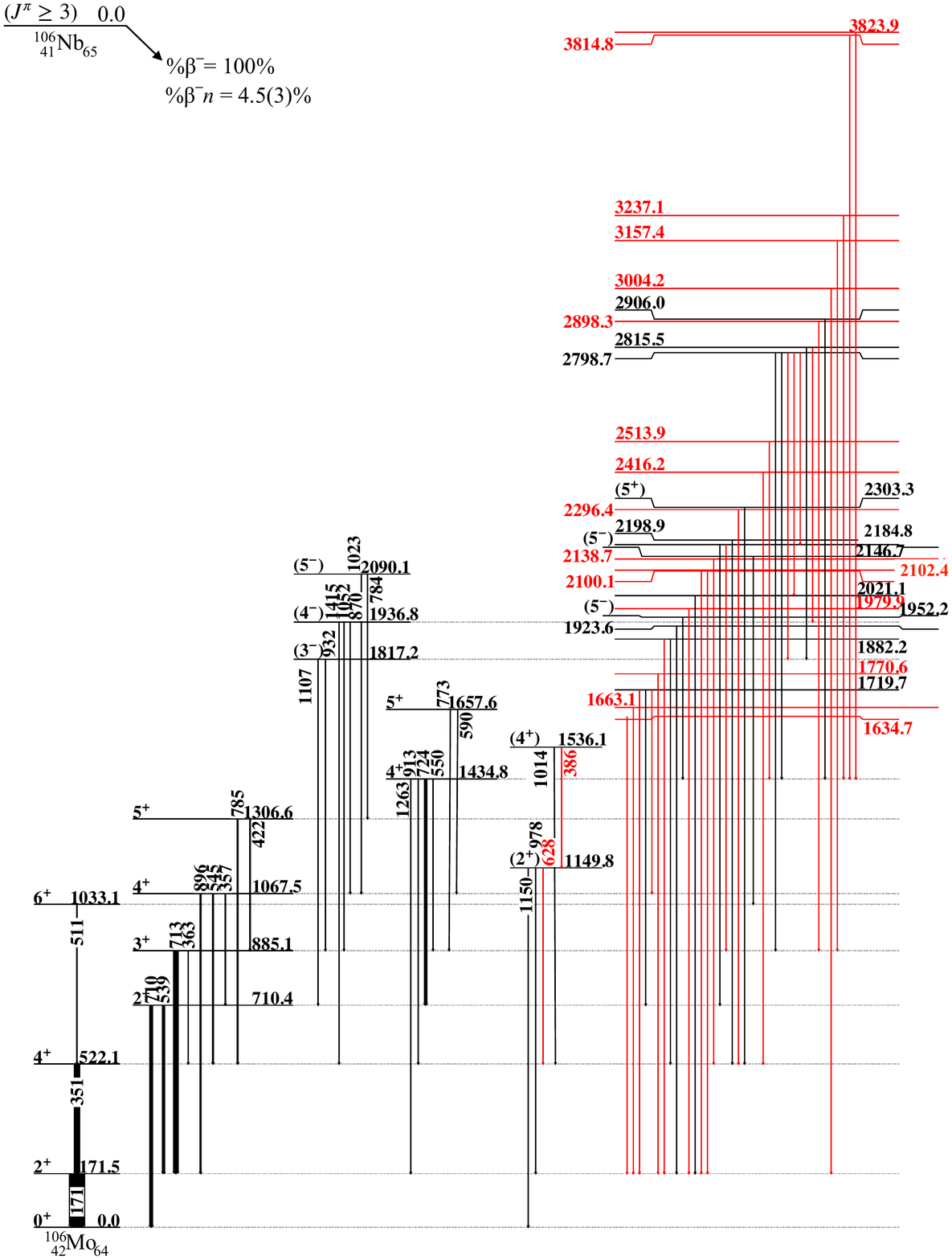}
\caption{\label{fig9} Proposed level scheme of \nuc{106}{Mo} following the $\beta$ decay of \nuc{106}{Nb}. New excited states and $\gamma$-ray transitions are in red. Spins and parities, and the $\beta$-delayed neutron emission value are adopted from Ref.~\cite{defrenne}.}
\end{figure*}

Shizuma $et~al$ \cite{shizuma} reported the location of the $yrast$ \jp{2}{+}{1}, \jp{4}{+}{1} and \jp{6}{+}{1} states, and identified candidates for the \jp{2}{+}{2}, \jp{3}{+}{1} and \jp{0}{+}{2} levels, while the work of Ha~$et~al$ \cite{ha} extended the level scheme up to $\approx$3~MeV. Here, we confirm the locations of 26 previously known excited states and 41 $\gamma$-ray transitions \cite{shizuma, ha}, and further expand the level scheme up to $\approx$4~MeV with an additional 16 excited states and 26 $\gamma$-ray transitions. In this manuscript, transitions and levels referred to as ``new'' are in relation to both Ref.~\cite{defrenne} and the recent observations reported in Ref.~\cite{ha}. The proposed expansion of the level scheme is provided in Fig.~\ref{fig9}. Fourteen of these excited states are associated with rotational-band structures identified in prompt spectroscopy of actinide fission fragments \cite{defrenne}. A summary of the excited states observed in this work is provided in Table~\ref{summary}, including level energies and spin-parity assignments, energies and branching ratios of depopulating transitions, and apparent $\beta$-feeding intensities. Where possible, $\gamma$-decay branching ratios for transitions depopulating each level have also been obtained by gating on a strong transition that feeds the level under inspection. Transition intensities reported in Ref.~\cite{ha} are provided for reference where they are available.

While the decay scheme has been extended extensively from Refs.~\cite{shizuma, ha}, the highest-lying level at $\approx$4~MeV is still $\approx$3~MeV below the neutron separation energy of 6.869~MeV \cite{defrenne}. Therefore, it is likely that a `Pandemonium' \cite{hardy} of direct $\beta$ feeding occurs to a high-density region of weakly populated states within this energy range. Such states are known to be beyond the sensitivity of discrete-line spectroscopy, and so further measurement of this nucleus adopting a technique such as `total absorption gamma-ray spectroscopy' will be required. For this reason, limits are quoted for the apparent $\beta$-feeding intensities. 

In this study, we confirm the locations of most excited states and transitions presented in Ref.~\cite{ha}. Four $\gamma$ rays were not observed: the 188-keV (2$^+_2$$\rightarrow$~4$^+_1$), 175-keV (3$^+_1$$\rightarrow$~2$^+_2$), 223-keV ($J$$^{\pi}$$\rightarrow$~5$^-$), and 1624-keV (5$^-$$\rightarrow$~4$^+$) transitions. Examples of gated spectra in which the low-energy transitions would be expected are presented in Fig.~\ref{fig8}. The 1624-keV $\gamma$ ray would be observed in the 351-keV gate of Fig.~\ref{fig6}. With the proposed 188-keV, 223-keV, and 1624-keV transitions, we do not observe a significant rise above fluctuations in the background at these energies. The 175-keV transition, if present, may be obscured by the dominant 172-keV transition. Reference~\cite{ha} lists a 1930-keV (2815~$\rightarrow$~885) transition; in this work, we only observe that $\gamma$ ray in coincidence with the 172-keV one and therefore, suggest a different placement in the level scheme with a new level at 2102~keV. 

We note two discrepancies with the low-lying states observed by Shizuma~\textit{et al} \cite{shizuma}: namely, the 957-keV (\jp{(0}{+}{2})) level and the 1280-keV one of unknown spin and parity. Tentative placement of the 957-keV level was based on the observation of a 785-keV $\gamma$ ray in coincidence with the 172-keV transition. The non-observation of a 957-keV $\gamma$ ray connecting this level to the ground state was suggested as evidence for this being the \jp{0}{+}{2} level. Two $\gamma$ rays with similar energies (784~keV and 785~keV) depopulating the 2090-keV and 1307-keV levels, respectively, were identified in prompt-fission studies. Coincidence relationships observed in the current work are consistent with this decay pattern, and confirmed by Ref.~\cite{ha}.

\clearpage
\setlength{\LTcapwidth}{17.8cm}
\begin{longtable*}{@{\extracolsep{0.8cm}}lcccccccc}
\caption{\label{summary} The $\gamma$-ray transitions and excited states in \nuc{106}{Mo} observed in this work following the $\beta$ decay of \nuc{106}{Nb}. Initial-level (E$_{\rm{i}}$), final-level (E$_{\rm{f}}$) and $\gamma$-ray (E$_{\gamma}$) energies are given in keV; uncertainties are discussed in the text. Spins and parities are from Ref.~\cite{defrenne} or proposed from the current work ($^{a}$). Transition intensities (I$_{\gamma}$) are normalized to the 172-keV transition (100(2) units). Transition intensities (I$_{\gamma}^{lit}$) and $\beta$-feeding intensities (I$_{\beta^-}^{lit}$) presented in Ref.~\cite{ha} are included here for comparison. Limitations of the apparent $\beta$-feeding intensities (I$_{\beta^-}$) from this work are discussed in the text. For absolute intensity per 100 parent decays multiply I$_{\gamma}$ by 0.71(8).} \\

\hline\hline \\[-0.2cm]
\multicolumn{1}{l}{$\rm{E_i}$} & 
\multicolumn{1}{c}{$\rm{J}^{\pi}_i$} & 
\multicolumn{1}{c}{$\rm{E}_{\gamma}$} & 
\multicolumn{1}{c}{$\rm{E_f}$} & 
\multicolumn{1}{c}{$\rm{J}^{\pi}_f$} & 
\multicolumn{1}{c}{$\rm{I}_{\gamma}$} & 
\multicolumn{1}{c}{$\rm{I}_{\gamma}^{lit}$} & 
\multicolumn{1}{c}{$\rm{I}_{\beta^-}$}   &
\multicolumn{1}{c}{$\rm{I}_{\beta^-}^{lit}$}   \\[0.1cm] 
\multicolumn{1}{l}{(keV)} & 
\multicolumn{1}{c}{} & 
\multicolumn{1}{c}{(keV)} & 
\multicolumn{1}{c}{(keV)} & 
\multicolumn{1}{c}{} & 
\multicolumn{1}{c}{($\%$)} & 
\multicolumn{1}{c}{($\%$)} & 
\multicolumn{1}{c}{($\%$)} & 
\multicolumn{1}{c}{($\%$)}  	\\[0.1cm]
\hline\hline  \\[-0.2cm]
\endfirsthead

\multicolumn{9}{c}%
{{\tablename\ \thetable{} -- continued}} \\
\hline\hline \\[-0.2cm]
\multicolumn{1}{l}{$\rm{E_i}$} & 
\multicolumn{1}{c}{$\rm{J}^{\pi}_i$} & 
\multicolumn{1}{c}{$\rm{E}_{\gamma}$} & 
\multicolumn{1}{c}{$\rm{E_f}$} & 
\multicolumn{1}{c}{$\rm{J}^{\pi}_f$} & 
\multicolumn{1}{c}{$\rm{I}_{\gamma}$} & 
\multicolumn{1}{c}{$\rm{I}_{\gamma}^{lit}$} & 
\multicolumn{1}{c}{$\rm{I}_{\beta^-}$}   &
\multicolumn{1}{c}{$\rm{I}_{\beta^-}^{lit}$}   \\[0.1cm] 
\multicolumn{1}{l}{(keV)} & 
\multicolumn{1}{c}{} & 
\multicolumn{1}{c}{(keV)} & 
\multicolumn{1}{c}{(keV)} & 
\multicolumn{1}{c}{} & 
\multicolumn{1}{c}{($\%$)} & 
\multicolumn{1}{c}{($\%$)} & 
\multicolumn{1}{c}{($\%$)} & 
\multicolumn{1}{c}{($\%$)}  	\\[0.1cm]
\hline\hline  \\[-0.2cm]
\endhead

 \\[-0.55cm]
\hline\hline  
\endfoot

0	&	0$^{+}$$^{}$	&	--	&	--	&	--	&	--	&	--	&	0	&	<8.4	\\ [0.15cm] \hline \\[-0.2cm]
171.49(9)	&	2$^{+}$$^{}$	&	171.5(1)	&	0	&	0$^{+}$$^{}$	&	100(2)	&	100.0(5)	&	10(3)	&	7.3(8)	\\ [0.15cm] \hline \\[-0.2cm]
522.08(11)	&	4$^{+}$$^{}$	&	350.6(1)	&	171.49(9)	&	2$^{+}$$^{}$	&	38.6(7)	&	43.8(5)	&	12.1(8)	&	9.1(14)	\\ [0.15cm] \hline \\[-0.2cm]
710.36(11)	&	2$^{+}$$^{}$	&	538.9(2)	&	171.49(9)	&	2$^{+}$$^{}$	&	13.6(4)	&	15.6(3)	&	6.8(8)	&	2.8(6)	\\ [0.1cm]
	&		&	710.3(2)	&	0	&	0$^{+}$$^{}$	&	15.7(4)	&	15.2(3)	&		&		\\ [0.15cm] \hline \\[-0.2cm]
885.07(12)	&	3$^{+}$$^{}$	&	363.0(4)	&	522.08(11)	&	4$^{+}$$^{}$	&	1.0(1)	&	0.7(2)	&	12.0(7)	&	8.7(7)	\\ [0.1cm]
	&		&	713.6(1)	&	171.49(9)	&	2$^{+}$$^{}$	&	29.2(6)	&	31.9(4)	&		&		\\ [0.15cm] \hline \\[-0.2cm]
1033.08(23)	&	6$^{+}$$^{}$	&	511.0(2)	&	522.08(11)	&	4$^{+}$$^{}$	&	2.5(2)	&	8.2(15)	&	1.6(1)	&	5.5(11)	\\ [0.15cm] \hline \\[-0.2cm]
1067.50(12)	&	4$^{+}$$^{}$	&	357.2(1)	&	710.36(11)	&	2$^{+}$$^{}$	&	1.7(2)	&	2.1(2)	&	6.1(4)	&	7.9(5)	\\ [0.1cm]
	&		&	545.4(2)	&	522.08(11)	&	4$^{+}$$^{}$	&	4.9(2)	&	7.6(2)	&		&		\\ [0.1cm]
	&		&	896.0(2)	&	171.49(9)	&	2$^{+}$$^{}$	&	5.8(2)	&	6.1(2)	&		&		\\ [0.15cm] \hline \\[-0.2cm]
1149.80(9)	&	(2$^{+}$)$^{}$	&	628.0(4)	&	522.08(11)	&	4$^{+}$$^{}$	&	0.7(1)	&		&	2.4(3)	&	1.6(2)	\\ [0.1cm]
	&		&	978.2(2)	&	171.49(9)	&	2$^{+}$$^{}$	&	2.1(2)	&	2.3(2)	&		&		\\ [0.1cm]
	&		&	1149.8(1)	&	0	&	0$^{+}$$^{}$	&	1.9(2)	&		&		&		\\ [0.15cm] \hline \\[-0.2cm]
1306.60(19)	&	5$^{+}$$^{}$	&	421.5(2)	&	885.07(12)	&	3$^{+}$$^{}$	&	1.9(1)	&	3.5(2)	&	3.9(2)	&	5.4(8)	\\ [0.1cm]
	&		&	784.7(5)	&	522.08(11)	&	4$^{+}$$^{}$	&	3.6(2)	&	5.5(7)	&		&		\\ [0.15cm] \hline \\[-0.2cm]
1434.78(12)	&	4$^{+}$$^{}$	&	549.8(2)	&	885.07(12)	&	3$^{+}$$^{}$	&	4.2(2)	&	6.9(2)	&	5.7(6)	&	7.0(5)	\\ [0.1cm]
	&		&	724.4(1)	&	710.36(11)	&	2$^{+}$$^{}$	&	12.1(6)	&	14.0(3)	&		&		\\ [0.1cm]
	&		&	912.7(1)	&	522.08(11)	&	4$^{+}$$^{}$	&	0.6(1)	&		&		&		\\ [0.1cm]
	&		&	1263.2(4)	&	171.49(9)	&	2$^{+}$$^{}$	&	1.5(1)	&	1.4(2)	&		&		\\ [0.15cm] \hline \\[-0.2cm]
1536.1(3)	&	(4$^{+}$)$^{}$	&	386.1(5)	&	1149.80(9)	&	(2$^{+}$)$^{}$	&	1.4(4)	&		&	2.0(3)	&	1.0(2)	\\ [0.1cm]
	&		&	1014.1(3)	&	522.08(11)	&	4$^{+}$$^{}$	&	1.4(1)	&	1.5(3)	&		&		\\ [0.15cm] \hline \\[-0.2cm]
1634.70(22)	&		&	1463.2(2)	&	171.49(9)	&	2$^{+}$$^{}$	&	0.4(1)	&		&	0.3(1)	&		\\ [0.15cm] \hline \\[-0.2cm]
1657.59(24)	&	5$^{+}$$^{}$	&	590.0(3)	&	1067.50(12)	&	4$^{+}$$^{}$	&	0.9(2)	&		&	1.4(2)	&	1.0(1)	\\ [0.1cm]
	&		&	772.6(3)	&	885.07(12)	&	3$^{+}$$^{}$	&	1.1(1)	&	1.4(2)	&		&		\\ [0.15cm] \hline \\[-0.2cm]
1663.10(22)	&		&	1491.6(2)	&	171.49(9)	&	2$^{+}$$^{}$	&	0.4(1)	&		&	0.3(1)	&		\\ [0.15cm] \hline \\[-0.2cm]
1719.75(16)	&		&	652.4(2)	&	1067.50(12)	&	4$^{+}$$^{}$	&	0.7(2)	&		&	1.7(2)	&	0.9(1)	\\ [0.1cm]
	&		&	1009.2(2)	&	710.36(11)	&	2$^{+}$$^{}$	&	1.2(2)	&	1.3(2)	&		&		\\ [0.1cm]
	&		&	1548.3(3)	&	171.49(9)	&	2$^{+}$$^{}$	&	0.5(1)	&		&		&		\\ [0.15cm] \hline \\[-0.2cm]
1770.6(4)	&		&	1599.1(4)	&	171.49(9)	&	2$^{+}$$^{}$	&	0.4(1)	&		&	0.3(1)	&		\\ [0.15cm] \hline \\[-0.2cm]
1817.26(23)	&	(3$^{-}$)$^{}$	&	932.2(3)	&	885.07(12)	&	3$^{+}$$^{}$	&	1.5(2)	&	2.0(2)	&	2.4(3)	&	4.9(4)	\\ [0.1cm]
	&		&	1106.9(4)	&	710.36(11)	&	2$^{+}$$^{}$	&	3.8(4)	&	7.4(3)	&		&		\\ [0.6cm] 
1882.15(21)	&		&	1359.7(5)	&	522.08(11)	&	4$^{+}$$^{}$	&	1.9(2)	&	2.9(2)	&	1.9(1)	&	2.0(2)	\\ [0.1cm]
	&		&	1710.7(2)	&	171.49(9)	&	2$^{+}$$^{}$	&	0.8(1)	&		&		&		\\ [0.19cm] \hline \\[-0.2cm]
1923.60(22)	&		&	1752.1(2)	&	171.49(9)	&	2$^{+}$$^{}$	&	1.0(1)	&	1.6(2)	&	0.7(1)	&	1.1(2)	\\ [0.15cm] \hline \\[-0.2cm]
1936.79(18)	&	(4$^{-}$)$^{}$	&	869.5(3)	&	1067.50(12)	&	4$^{+}$$^{}$	&	1.5(3)	&	2.0(2)	&	2.0(3)	&	3.5(3)	\\ [0.1cm]
	&		&	1051.6(2)	&	885.07(12)	&	3$^{+}$$^{}$	&	2.3(2)	&	3.1(2)	&		&		\\ [0.1cm]
	&		&	1414.5(4)	&	522.08(11)	&	4$^{+}$$^{}$	&	0.5(1)	&		&		&		\\ [0.15cm] \hline \\[-0.2cm]
1952.18(23)	&	(5$^{-}$)$^{}$	&	517.4(2)	&	1434.78(12)	&	4$^{+}$$^{}$	&	3.8(3)	&	4.6(2)	&	2.7(2)	&	2.3(2)	\\ [0.15cm] \hline \\[-0.2cm]
1979.90(22)	&		&	1808.4(2)	&	171.49(9)	&	2$^{+}$$^{}$	&	0.7(1)	&		&	0.5(1)	&		\\ [0.15cm] \hline \\[-0.2cm]
2021.1(3)	&	(3,4$^{}$)$^{a}$	&	1849.5(4)	&	171.49(9)	&	2$^{+}$$^{}$	&	3.2(2)	&	4.1(3)	&	1.9(2)	&	2.9(3)	\\ [0.15cm] \hline \\[-0.2cm]
2090.11(20)	&	(5$^{-}$)$^{}$	&	783.5(2)	&	1306.60(19)	&	5$^{+}$$^{}$	&	0.2(1)	&	1.3(7)	&	0.6(1)	&	2.7(5)	\\ [0.1cm]
	&		&	1022.6(2)	&	1067.50(12)	&	4$^{+}$$^{}$	&	0.6(2)	&	2.5(2)	&		&		\\ [0.15cm] \hline \\[-0.2cm]
2100.1(4)	&		&	1928.6(4)	&	171.49(9)	&	2$^{+}$$^{}$	&	1.2(3)	&	2.7(2)	&	0.9(2)	&		\\ [0.15cm] \hline \\[-0.2cm]
2102.4(4)	&		&	1930.9(4)	&	171.49(9)	&	2$^{+}$$^{}$	&	1.4(3)	&		&	1.0(2)	&		\\ [0.15cm] \hline \\[-0.2cm]
2138.7(4)	&	(4,5$^{}$)$^{a}$	&	1616.6(3)	&	522.08(11)	&	4$^{+}$$^{}$	&	1.4(1)	&		&	1.0(1)	&		\\ [0.15cm] \hline \\[-0.2cm]
2146.7(8)	&	(5$^{-}$)$^{}$	&	1113.6(7)	&	1033.08(23)	&	6$^{+}$$^{}$	&	0.3(1)	&	0.3(2)	&	0.18(5)	&	0.5(2)	\\ [0.15cm] \hline \\[-0.2cm]
2184.78(20)	&	(3,4$^{}$)$^{a}$	&	1299.9(3)	&	885.07(12)	&	3$^{+}$$^{}$	&	0.4(1)	&		&	0.5(2)	&	0.9(1)	\\ [0.1cm]
	&		&	1474.4(3)	&	710.36(11)	&	2$^{+}$$^{}$	&	0.9(3)	&	1.3(2)	&		&		\\ [0.15cm] \hline \\[-0.2cm]
2198.9(4)	&	(4,5$^{}$)$^{a}$	&	1676.8(3)	&	522.08(11)	&	4$^{+}$$^{}$	&	1.3(1)	&	2.2(2)	&	0.9(1)	&	1.5(2)	\\ [0.15cm] \hline \\[-0.2cm]
2296.4(6)	&	(4,5$^{}$)$^{a}$	&	1774.3(5)	&	522.08(11)	&	4$^{+}$$^{}$	&	0.9(1)	&		&	0.6(1)	&		\\ [0.15cm] \hline \\[-0.2cm]
2303.3(4)	&	(5$^{+}$)$^{}$	&	1781.2(3)	&	522.08(11)	&	4$^{+}$$^{}$	&	0.7(1)	&	1.4(2)	&	0.5(1)	&	1.0(1)	\\ [0.15cm] \hline \\[-0.2cm]
2416.2(4)	&	(4,5$^{}$)$^{a}$	&	1894.1(3)	&	522.08(11)	&	4$^{+}$$^{}$	&	0.5(1)	&		&	0.4(1)	&		\\ [0.15cm] \hline \\[-0.2cm]
2513.9(4)	&	(4,5$^{}$)$^{a}$	&	1079.1(3)	&	1434.78(12)	&	4$^{+}$$^{}$	&	0.5(1)	&		&	0.4(1)	&		\\ [0.15cm] \hline \\[-0.2cm]
2798.70(19)	&	(4$^{-}$)$^{a}$	&	614.0(2)	&	2184.78(20)	&	(3,4)$^{}$$^{a}$	&	0.6(1)	&		&	4.1(3)	&	5.2(3)	\\ [0.1cm]
	&		&	777.5(4)	&	2021.1(3)	&	(3,4)$^{}$$^{a}$	&	0.6(1)	&		&		&		\\ [0.1cm]
	&		&	981.1(5)	&	1817.26(23)	&	(3$^{-}$)$^{}$	&	0.5(1)	&		&		&		\\ [0.1cm]
	&		&	1363.9(3)	&	1434.78(12)	&	4$^{+}$$^{}$	&	3.1(3)	&	5.9(2)	&		&		\\ [0.1cm]
	&		&	1913.6(3)	&	885.07(12)	&	3$^{+}$$^{}$	&	0.9(2)	&	1.5(1)	&		&		\\ [0.15cm] \hline \\[-0.2cm]
2815.5(3)	&		&	878.6(3)	&	1936.79(18)	&	(4$^{-}$)$^{}$	&	1.4(2)	&		&	2.0(3)	&	3.5(2)	\\ [0.1cm]
	&		&	998.5(4)	&	1817.26(23)	&	(3$^{-}$)$^{}$	&	1.4(3)	&	2.3(1)	&		&		\\ [0.15cm] \hline \\[-0.2cm]
2898.3(5)	&		&	2013.2(4)	&	885.07(12)	&	3$^{+}$$^{}$	&	0.5(1)	&		&	0.4(1)	&		\\ [0.15cm] \hline \\[-0.2cm]
2906.0(6)	&	(4,5$^{}$)$^{a}$	&	1471.2(5)	&	1434.78(12)	&	4$^{+}$$^{}$	&	1.4(2)	&	1.7(2)	&	1.0(1)	&	1.2(2)	\\ [0.15cm] \hline \\[-0.2cm]
3004.2(4)	&		&	2832.7(4)	&	171.49(9)	&	2$^{+}$$^{}$	&	1.2(2)	&		&	0.8(1)	&		\\ [0.15cm] \hline \\[-0.2cm]
3157.4(5)	&		&	2272.3(4)	&	885.07(12)	&	3$^{+}$$^{}$	&	0.5(1)	&		&	0.4(1)	&		\\ [0.15cm] \hline \\[-0.2cm]
3237.1(7)	&	(4,5$^{}$)$^{a}$	&	1802.3(7)	&	1434.78(12)	&	4$^{+}$$^{}$	&	0.4(1)	&		&	0.3(1)	&		\\ [0.3cm] 
3814.8(6)	&	(4,5$^{}$)$^{a}$	&	2380.0(5)	&	1434.78(12)	&	4$^{+}$$^{}$	&	0.4(2)	&		&	0.3(1)	&		\\ [0.15cm] \hline \\[-0.2cm]
3823.9(5)	&	(4,5$^{}$)$^{a}$	&	2389.1(4)	&	1434.78(12)	&	4$^{+}$$^{}$	&	0.8(2)	&		&	0.5(1)	&		\\ [-0.1cm]

\label{table:levels}
\end{longtable*}

While the location of the \jp{0}{+}{2} state is certainly not at 957~keV, several candidates are described below. However, further experiments are necessary to confirm the location and nature of these levels. Similarly, the 1280-keV level was suggested on the basis of an 1108-keV $\gamma$-ray transition also found to be in coincidence with the 172-keV one. Our analysis instead supports the placement of the 1108-keV transition as connecting the (3$^-$) state at 1817~keV to the 2$^+$ state at 710~keV. The repositioning of this $\gamma$-ray transition is also noted in Ref.~\cite{ha}, so there is no excited state at 1280~keV.

\subsubsection{\textbf{Confirmation of known states}}

The 2$^+_g$, 4$^+_g$, and 6$^+_g$ members of the $yrast$ rotational band built on a prolate-deformed 0$^+$ ground state ($g$) have been identified. While the locations of the 8$^+_g$ and 10$^+_g$ members are known \cite{defrenne}, they are not fed by $\beta$-decay. The band built on the $K^{\pi}$~=~2$^+$ ($\gamma$ band), 710-keV level is observed up to the 5$^+_{\gamma}$ member at 1307~keV. 
 
Intra-band, $\Delta J$~=~2 transitions (4$^+_{\gamma}$$\rightarrow$~2$^+_{\gamma}$ and 5$^+_{\gamma}$$\rightarrow$~3$^+_{\gamma}$) were identified, however there was no evidence for $\Delta J$~=~1 transitions between the band levels. Known inter-band transitions between the $\gamma$ and ground-state bands were observed, with the exception of the spin-increasing 5$^+_{\gamma}$$\rightarrow$~6$^+_g$ one. Branching ratios measured in the current work indicate that the 2$^+_{\gamma}$$\rightarrow$~0$^+_g$ decay path is slightly enhanced with respect to the 2$^+_{\gamma}$$\rightarrow$~2$^+_g$ transition. 

The strongest $\gamma$ ray observed to feed the $K^{\pi}$~=~2$^+$ bandhead is the 724-keV transition from the $K$~=~4, 1435-keV level. Guessous~\textit{et al} identified this as a candidate double-phonon $\gamma$-vibrational state \cite{guessous}. The known 5$^+$ member of this band is also identified in the current work, although the 223-keV transition between these two levels was not observed. Three levels corresponding to a $K^{\pi}$~=~3$^-$, negative-parity band, suggested to arise from a $\nu\frac{3}{2}$[411]$\otimes$$\nu\frac{3}{2}$[532] configuration \cite{defrenne}, have been identified in this work. The $\gamma$ rays connecting each of the levels in this sequence to the $\gamma$-vibrational band were observed. Two levels associated with a proposed $K^{\pi}$~=~(2$^+$) band were also identified at 1150~keV and 1536~keV. Bandheads of the three other two-quasiparticle structures listed in the adopted levels have been observed: the (5$^-$), 1952-keV level ($\nu\frac{5}{2}$[413]$\otimes$$\nu\frac{5}{2}$[532]); the (5$^-$), 2147-keV state ($\pi\frac{7}{2}$[413]$\otimes$$\pi\frac{3}{2}$[301]); and the (5$^+$), 2302-keV level ($\pi\frac{1}{2}$[420]$\otimes$$\pi\frac{9}{2}$[404]). A single $\gamma$ ray was observed to depopulate each of these states; any other depopulating transitions that may occur fall below the level of sensitivity, $I_{\gamma}$~$\geq$~0.02$\times I_{172}$, of the present measurement. 

\subsubsection{\textbf{Identification of new states}}

Seventeen previously unobserved excited states have been added in this work: ten decay directly by single transitions to levels within the yrast band, three are connected to the $\gamma$ band, and four are connected to the proposed harmonic, two-phonon $\gamma$-vibrational state \cite{guessous}. While it is not possible to assign firm spins and parities to these new levels with the current data, it was possible to place spin constraints on some from the observed decay pattern. Where available, these are described in the text. Spin-parity assignments listed in Table~\ref{summary} without parentheses are taken from the literature \cite{defrenne}. 

Nine excited states are each observed to have a single $\gamma$-decay branch that connects it to one of the levels with a firm 4$^+$ assignment. The weak apparent $\beta$-feeding intensities and lack of $\gamma$-decay branches to 2$^+$ or 3$^+$ states suggest these are of moderate spin, and so a $J~=~(4)$~or~(5) assignment is suggested for these levels. The excited state at 2799 keV is unusual in that the apparent $\beta$-feeding intensity is larger than that of any other state observed above 2-MeV excitation energy, and multiple $\gamma$-decay pathways from the state were identified. Strong feeding to the 1435-keV, 4$^+$ level and two $J~=~3$ levels and relatively low log-$ft$ value of 6.07(1) suggest a tentative \jp{(4}{-}{}) assignment is appropriate for this level. 


\section{DISCUSSION AND CONCLUSIONS}\label{sec:discussion}

The neutron-rich nuclei at $A$~$\approx$~100 have proven to be technically challenging from both experimental and theoretical points of view. Ground-state charge-radii measurements point to a rapid spherical-to-prolate-deformed shape transition between $N$~=~58 and $N$~=~60 \cite{rodriguez-guzman} similar to the well-established phenomenon observed between stable $N$~=~88 and $N$~=~90 rare-earth nuclei \cite{casten}. This phenomenon appears to be strongest in zirconium ($Z$~=~40) \cite{campbell}, persists in neighbouring strontium ($Z$~=~38) \cite{buchinger} and weakens in molybdenum ($Z$~=~42) \cite{charlwood}, an effect attributed to the triaxial nature of the latter isotopes. This is supported by local trends in $E(2^+_1)$ and $B$($E$2; 0$^+_2$$\rightarrow$~2$^+_1$) values \cite{pritychenko}. 

Coulomb-excitation measurements with radioactive-ion beams \cite{gorgen, clement1} indicate that shape coexistence is prevalent in the region \cite{heyde}, whereby deformed \jp{0}{+}{2} states at $N$~$<$~60 migrate to become the ground states at $N$~$\geq$~60. Quantum phase transitions have been attributed as the driving force behind this rapid evolution of the nuclear shape \cite{togashi, kremer}. Beyond $N$~=~60, there is increasing evidence that the deformation softens towards the neutron drip-line and that the triaxial degree of freedom plays in an important role in the behaviour of neutron-rich molybdenum isotopes \cite{urban1, smith1, ding, watanabe, smith2, snyder, ralet}. 

The picture becomes more complex in the adjacent, odd-$Z$ niobium ($Z$~=~41) isotopes. In the case of \nuc{106}{Nb} ($N$~=~65), only a single investigation into the level scheme exists in the literature from prompt-fission spectroscopy \cite{luo}; direct observation of the $\beta$-decay properties of this nuclide are similarly rare. Initial observation of strong $\beta$-decay feeding to $J = 4, 5$ excited states in \nuc{106}{Mo} prompted further investigation. Lighter-mass, odd-odd Nb isotopes exhibit an alternating pattern of low-spin/high-spin $\beta$-decaying ground states and isomers. At \nuc{106}Nb, the traditional $N$~=~64 neutron sub-shell closure is crossed, exposing a new valence space. While it is unlikely that the pattern of $\beta$-decaying isomers (see above) continues into \nuc{106}{Nb}, it could explain the observed pattern in the $\gamma$-decay measurement.   

As discussed above, the new results indicate that the ground-state spin-parity assignment to \nuc{106}{Nb} should be revised. The adopted assignment, \jp{(1}{-}{}), of Ref.~\cite{defrenne} is based upon potential-energy surface (PES) and projected shell-model (PSM) calculations presented in Ref.~\cite{luo}. They predict a triaxial $\pi \frac{3}{2}^- [301] \bigotimes \nu \frac{5}{2}^+ [413]$ ground state with ($\beta$,$\gamma$)~=~(0.35,15$^{\circ}$) deformation parameters. At ($Z,N$) = (41,65), \nuc{106}{Nb} lies a long way from the single stable isotope, \nuc{93}{Nb}. Naively, one might predict the ground-state configuration to be dominated by a two-quasiparticle coupling of the odd proton and neutron outside the $Z=40$ and $N=64$ sub-shell closures, respectively. The works of Kurpeta $et~al.$ \cite{kurpeta} and Urban $et~al.$ \cite{urban2} provide the most-recent considerations of the neighbouring isotope, \nuc{107}{Nb}, and its isobar, \nuc{107}{Mo}. They suggest (5/2$^+$) and 1/2$^+$ ground states, respectively, for these nuclides from a combination of $\beta$-decay feeding and assessment of systematic trends. A prolate $\pi \frac{5}{2}^+ [422] \bigotimes \nu \frac{1}{2}^+ [411]$ configuration with ($\beta$,$\gamma$)~=~(0.32,0) was predicted for \nuc{106}{Nb} in the PES calculations of Ref.~\cite{luo}, however the excitation energy is 597 keV. With maximal spin coupling, as per the Gallagher-Moszkowski coupling rule \cite{gallagher}, a favoured 3$^+$ assignment would be expected. A 3$^+$ ground state could explain most of the $\beta$-decay feeding pattern observed in this work; the feeding to 3$^{\pm}$, and 4$^{\pm}$ states would then be accessible from allowed and first-forbidden $\beta$ decays. 

The observed feeding to 5$^{\pm}$ states would favour a \jp{(4}{\pm}{}) assignment. Maximal spin coupling of the $\pi \frac{3}{2}^- [301] \bigotimes \nu \frac{5}{2}^+ [413]$ configuration from Ref.~\cite{luo} discussed above would result in a \jp{4}{-}{} ground state; this assignment would violate the Gallagher-Moszkowski rule \cite{gallagher}. The requirement of such highly forbidden $\beta$ decays to explain the observed feeding from a supposed 1$^-$ ground state cannot be ignored. In light of our decay study, non-observation of a $\beta$-decaying isomer from our mass measurement, and the recent work of Ha~$et~al$ \cite{ha}, it is clear that the assumption of a \jp{1}{-}{} ground state is incorrect and the spin assignments of all excited states in \nuc{106}{Nb} are in need of a full reappraisal.

If the $^{106}$Nb ground-state spin and parity were $J=3^+$, any $\beta$ decay to the $^{106}$Mo ground state is $\Delta J=3$, $\Delta \pi =0$. This would be a unique, second-forbidden decay. In nature, 12 such cases are documented \cite{singh2}, with the minimum log-$ft$ being 13.9. With our new mass and decay half-life measurements, this would correspond to a branch of $<$10$^{-6}$ $\%$ -- far below the experimental sensitivity and sufficiently close to zero to not influence the calculated distribution of strength or normalization. If the spin and parity of $^{106}$Nb is $J=4^-$, the ground-state $\beta$ decay is unique, third forbidden. The only documented example of such a decay in the periodic table has a log-$ft$ value of 21, implying that the branch is $<$$10^{-11}$ $\%$.

While the interpretation of \nuc{106}{Nb} is uncertain, the picture is much clearer for \nuc{106}{Mo}. Several theoretical studies \cite{skalski, zhang, xiang, abusara, nomura} point to an emergence of triaxial softness in the neutron-rich molybdenum isotopes beyond $N=60$. In each case, triaxiality is essential to reproduce experimental observations. This undoubtedly contributes to the evolution of collectivity across the isotopic chain.

The distribution of excited states in \nuc{106}{Mo} directly fed by the $\beta$ decay of \nuc{106}{Nb} has been mapped up to $\approx$~4~MeV. A gradual, somewhat linear, increase in cumulated $\beta$-feeding strength is observed between 1~MeV and 2~MeV. An appreciable difference exists from the pattern of feeding to low-lying states reported in Ref.~\cite{defrenne}. Reference~\cite{ha} reports an upper limit of 8.4~$\%$ direct feeding to the ground state; a 4$^-$$\rightarrow$0$^+$ $\beta$ transition most certainly would not be observed with such a large intensity, or short decay half-life. While the possibility of a unique first-forbidden decay (4$^-$$\rightarrow$2$^+$) cannot be excluded by the log-$ft$ values, the large intensity ($<12.7\%$) is unusual for such a decay mode. Large feeding intensities that result from suggested unique first-forbidden $\beta$ decay have also been reported in neighbouring \nuc{108,110}{Mo} \cite{ha}. However, the apparent feeding intensities are also susceptible to strong Pandemonium effects, discussed above.  

Several of the new excited states observed in this work may be considered candidates for the elusive first-excited \jp{0}{+}{} state. If the \nuc{106}{Nb} ground state has a $J$~$\geq$~3 assignment, as expected, the candidate \jp{0}{+}{} state would not be fed directly from $\beta$-decay. Shape coexistence appears to be well established in the region and, therefore, one would expect to observe a low-lying \jp{0}{+}{} excited state in \nuc{106}{Mo}; excited \jp{0}{+}{2} states in \nuc{108,110}{Mo} are reported at 893.4~keV and 1042.2~keV, respectively, in Ref.~\cite{ha}. Of the 17 new levels in \nuc{106}{Mo}, seven are observed to decay via a single transition to the \jp{2}{+}{1} state. The present data are sensitive to $\gamma$ rays with intensities of $\approx$0.2$\%$ relative to the 172-keV transition. While the possibility of weak ground-state feeding or branches to other states below this level of sensitivity cannot be ruled out, determining the true nature and location of any \jp{0}{+}{} levels will require dedicated experimental searches. A search for mono-energetic $E$0 electrons from the direct decay of the \jp{0}{+}{2} level to the ground state might be productive; this would be the preferred decay mode if the co-existence is strong and the \jp{0}{+}{2} state lies only tens of keV above the \jp{2}{+}{1} level.

In the $A~\approx$~100 neutron-rich nuclei, despite very large deformation, $K$-isomers have not been found, possibly due to the fragility of the shell-stabilised shapes. In this specific case, the combination of a high $Q$-value for \nuc{106}{Nb} $\beta$ decay and soft shapes in the decay product leads to unusually large fragmentation, both in $\beta$-decay strength and the subsequent $\gamma$-decay cascade. This, then, appears to be a situation where `Pandemonium' must occur, and so inferring the population of individual states from the observed $\gamma$ intensity balance becomes problematic. Inferring log-$ft$ values, and thus spin assignments and structure information, from these $\beta$-decay branches, as suggested by Ha~$et~al$ \cite{ha}, may be optimistic.

In summary, ground-state and $\beta$-decay properties of the very-neutron-rich nuclide \nuc{106}{Nb} have been studied at the CARIBU facility at Argonne National Laboratory. The ground-state mass of \nuc{106}{Nb} was measured to be $-$66202.0(13)~keV with the Canadian Penning Trap, which is consistent with the 2016 Atomic Mass Evaluation. This work ruled out the existence of a long-lived, high-spin, $\beta$-decaying isomer above $\approx$5~keV excitation in \nuc{106}{Nb}. Detailed $\beta$-delayed $\gamma$-ray spectroscopy of the progeny, \nuc{106}{Mo}, was performed with the X-Array and SATURN low-energy decay-spectroscopy station. The $\beta$-decay half-life was found to be T$_{1/2}$~=~1.097(21)~s. The decay scheme of \nuc{106}{Mo} has been extended up to $\approx$4~MeV. The combination of enhanced apparent $\beta$-feeding intensity to $J$~=~3-5 states in \nuc{106}{Mo}, and non-observation of a $\beta$-decaying isomer, leads to the conclusion that the ground-state spin-parity assignment for \nuc{106}{Nb}, and those of excited states in this nuclide, should be reassessed.

In future measurements with the X-Array, the addition of the MR-TOF separator to the CARIBU low-energy beam line and development of a new low-background, low-energy experimental hall will greatly improve the beam purity and sensitivity of decay-spectroscopy experiments. This work highlights the pressing need for considerable theoretical effort to enable accurate interpretation of spectroscopic data obtained for very-neutron-rich exotic niobium isotopes. \\


\section{ACKNOWLEDGEMENTS}\label{sec:acknowledgements}

The authors wish to acknowledge the excellent work of the Physics Support group of the ATLAS Facility at Argonne National Laboratory. This material is based upon work supported by the Australian Research Council Discovery Project 120104176 (ANU), the U.S. Department of Energy, Office of Science, Office of Nuclear Physics under Grants No.~DE-FG02-94ER40848 (UML), No.~DEFG02-97ER41041 (UNC), and No.~DE-FG02-97ER41033 (TUNL), and Contract No.~DE-AC02-06CH11357 (ANL), the National Nuclear Security Administration, Office of Defense Nuclear Nonproliferation R\&D (NA-22) and NSERC (Canada) under Contract No. SAPPJ-2015-00034. This research used resources of ANL's ATLAS facility, which is a DOE Office of Science User Facility. Fig.~\ref{fig9} in this article has been created using the LevelScheme scientific figure preparation system [M. A. Caprio, Comput. Phys. Commun. 171, 107 (2005), \url{http://scidraw.nd.edu/levelscheme}]. \\


%


\end{document}